\documentstyle[12pt]{article}

\catcode`@=11
\def\@cite#1#2{$^{\scriptscriptstyle
\mbox{\rm\scriptsize#1\if@tempswa , #2\fi}}$}

\def\mite{\@ifnextchar [{\@tempswatrue\@mitex}{\@tempswafalse\@mitex[]}}

\def\@mitex[#1]#2{\if@filesw\immediate\write\@auxout{\string\citation{#2}}\fi
  \def\@mitea{}\@mite{\@for\@miteb:=#2\do
    {\@mitea\def\@mitea{,}\@ifundefined
       {b@\@miteb}{{\bf ?}\@warning
       {Red alert: `\@miteb' found on deck \thepage \space }}
\hbox{\csname b@\@miteb\endcsname}}}{#1}}

\def\@mite#1#2{$\mbox{\rm#1\if@tempswa , #2\fi}$}
\catcode`@=12

\def\onward{\addtocounter{section}{1} \setcounter{equation}{0} }

\newcommand{\eq}{\begin{equation}}
\newcommand{\en}{\end{equation}}
\newcommand{\ie}{{\it i.e.}}

\def\cH{{\cal H}}

\def\cO{{\cal O}}
\def\cL{{\cal L}}
\def\cJ{{\cal J}}
\def\cV{{\cal V}}

\def\half{{\frac{1}{2}}}
\def\sshalf{{\sst {1 \over 2}} }
\def\shalf{{\sct {1 \over 2}} }

\def\sst{\scriptscriptstyle}
\def\sct{\scriptstyle}
\def\ds{\displaystyle}
\def\VEV#1{\left\langle #1\right\rangle}
\def\vev#1{\langle #1 \rangle}

\def\tr{{\rm tr}}
\def\Tr{{\rm Tr}}
\def\ie{{\it i.e.~}}

\def\bfpi{\mbox{\boldmath $\pi$}}
\def\bftau{\mbox{\boldmath $\tau$}}
\def\bgcup{\overline{g}_{\pi \sN \sN}}
\def\gcup{g_{\pi \sN \sN}}
\def\sN{\sct N}
\def\Dn{D_{n}}
\def\su3{{\rm SU}(3)}
\def\u3{{\rm U}(3)}
\def\vr{\bfvec{r}}
\def\Nc{N_{c}}
\def\bfvec#1{{\bf #1}}

\def\rI{{\rm I}}

\begin{document}
\setlength{\unitlength}{0.25cm}
\thispagestyle{empty}

\hfill             {\sc BRX-TH}-338
\vspace{1.5cm}

\begin{center}
{\large\bf  CP-Violating Yukawa Couplings in the Skyrme Model  \\[0.25cm]
          and the Neutron Electric Dipole Moment\footnote{Supported
           in part by  the DOE under grant
               DE-FG02-92ER40706}}\\[2.0cm]

{\large      Harold A. Riggs and Howard J. Schnitzer }

\vspace{1.0cm}
{ \sl
\begin{tabular}{c}
   Department of Physics, Brandeis University \\
   Waltham, MA 02254          \\[0.4cm]
   {\rm November 1992}
\end{tabular}  }

\end{center}

\vfill
\begin{center}
{\sc Abstract}
\end{center}

\begin{quotation}
We argue that the large-$\Nc$ behaviour of the Yukawa couplings in
the Skyrme model involves issues more subtle than the vanishing of
linear fluctuations needed for classical stability of the skyrmion.
The chiral fluctuations about the skyrmion must be quantized in order to
reach a conclusion.  An improved quantization
procedure allows us to confront this question
directly.

The pion-nucleon coupling constants $\gcup$ (CP conserving) and
$\bgcup$ (CP violating) are calculated in the large-$\Nc$,
three-flavour Skyrme model by direct evaluation of the
leading matrix elements appearing in
the LSZ reduction formula. We find that $\gcup \sim \Nc^{{3 \over 2}}$,
but that, at most,
$\bgcup \sim m^2_\pi \Nc^{-\shalf}$.
These results
show that the leading contribution to the neutron
electric dipole moment in large-$\Nc$
Skyrme model is the direct one ($\Dn \sim \Nc m^2_\pi$), rather
than the pion loop contribution.
\end{quotation}
\vfill

\setcounter{page}{0}
\newpage
\setcounter{page}{1}

\vspace{1.0cm}
\noindent {\bf 1. Introduction}
\vspace{0.3cm}
\setcounter{section}{1}
\setcounter{footnote}{1}

Some time ago one of us considered chiral fluctuations
about a rotating, three-flavour skyrmion\cite{hjs1}
\eq
    U(\bfvec{r} ,t) ~=
 ~U_{\phi}(\bfvec{r} ,t) ~A(t) ~U_{0}(\vr ~\!) ~A^{-1}(t)
                                                ~U_{\phi}(\bfvec{r} ,t)
\label{rotsk}
\en
where $U_{0}(\bfvec{r})$ is a static solution of the $\u3$ Skyrme
model, $A(t)$ is an $\su3$ matrix of collective coordinates,   and
the matrix field (summation convention throughout)
\eq
    U_{\phi} = \exp\{ {i \over f_{\pi} }
    \phi_\alpha \lambda_\alpha         \}
\label{matf}
\en
is built from the nonet of pseudoscalar meson fields
$\phi_{\alpha}$ ($\alpha = 0, 1,  \ldots, 8$) and the Gell-Mann
matrices $\lambda_{a}$ ($a= 1, \ldots, 8$) with
$\lambda_0 = \sqrt{2/3} ~\!\rI$. Insertion of
expression~\ref{rotsk} into the Skyrme lagrangian, $\cL_{S}$,
yields the elegant result
\eq
    \cL_{S}( U(\vr,t)) = \cL_{S} (A U_{0} A^{-1}) +
                         \cL_{\sigma} (U_{2\phi}) +
                         \cL_{I}(A U_{0} A^{-1}; \phi) ~,
\label{Linsert}
\en
in which the right-hand side is isomorphic to the low-energy
soft-meson and baryon effective lagrangian
\eq
    \cL_{{\rm eff}} = \cL_{B}(\psi) + \cL_{\sigma}(\pi)+\cL_{I}(\psi;\pi)
\label{effL}
\en
which describes free baryons  interacting with the mesons of
the non-linear $\sigma$-model.
Since~\ref{effL} obeys all the soft-meson
theorems of chiral symmetry, one expects that these theorems
are obeyed by the skyrmion-meson system of~\ref{Linsert} as well.

The Skyrme lagrangian gives a {\it formal} result
for the Yukawa couplings of the skyrmion-meson interaction.
However, to arrive at a conclusive result for the Yukawa couplings
involves a number of subtleties which lead to the central
issues of this paper. Given that~\ref{Linsert} includes appropriate
CP violating terms, one can compute both the CP violating
pion-nucleon scalar coupling constant $\bgcup$ and the
CP conserving pseudoscalar coupling constant $\gcup$ appearing
in the interaction term of eq.~\ref{effL}
\eq
      \cL_{I}(\psi;\pi) =
\bfpi \cdot
\overline{\psi} \bftau
(i \gamma_5  \gcup + \bgcup ) \psi   ~.
\label{new}
\en
Since the Skyrme model is formulated in terms of both
chiral symmetry and the large $\Nc$ (number of colors) expansion,
one expects that its predictions for $\gcup$ and
$\bgcup$ are consistent with QCD in the same limits.
Na\"{\i}vely, the dependence of the couplings
on the pion (nucleon) mass $m_{\pi}$ ($M_{N}$) and the
number of colors $\Nc$
is, from~\ref{Linsert} and ref.~\mite{hjs1},
\eq
   \begin{array}{ccc}
  \gcup \sim \Nc^{3\over 2} &
  ~~~\bgcup \sim m_{\pi}^2 \Nc^{\sshalf}~~~ &
  M_{N} \sim \Nc   ~.
    \end{array}
\label{Ndepend}
\en
Current algebra provides an ostensibly model-independent
derivation of the leading contribution to the electric dipole moment
of the neutron $\Dn$ in the chiral limit\cite{dnchiral}
\eq
    \Dn = {\gcup ~\bgcup \over 4 \pi^2 M_{N}} \ln ({M_{N} \over m_{\pi}}) ~.
\label{domin}
\en
If eq.~\ref{Ndepend} were correct, the leading contribution
in the combined chiral and large $\Nc$ limit
(and so dominant prediction of the Skyrme model in these limits) would be
$\Dn \sim \Nc m_{\pi}^2 \ln m_{\pi}$.

Recently significant progress in resolving this
issue has been made in calculating $\gcup$ (in the two flavour
Skyrme model) by the authors of
ref.~\mite{hayashi2}.  It is the purpose of this paper to
calculate the CP conserving and and
CP violating $\pi$-N couplings  $\gcup$, and $\bgcup$ by the
methods of ref.~\mite{hayashi2}, generalized to the
three-flavour Skyrme model.  We will show that
although the pseudoscalar coupling $\gcup$
indeed has the leading behaviour of eq.~\ref{Ndepend}, the
CP violating coupling $\bgcup$ is down by a factor of at
least $\Nc^{-1}$ from that of eq.~\ref{Ndepend}. Therefore,
the pion contribution to $\Dn$ is at most
\eq
        \Dn \sim m^2_\pi \ln m_\pi
\en
and is therefore subdominant to the direct CP violating behaviour
$\Dn \sim m^2_\pi \Nc$ computed in ref.~\mite{dixon}.
(For a review see ref.~\mite{nir}.)

It is important to understand why there is an issue with
regard to the Skyrme model Yukawa couplings. These
couplings embody the linear fluctuations about the
Skyrme classical solution. For constant collective coordinates
($A={\rm ~constant}$), this linear coupling must vanish
for the classical solution to be stable under linear
fluctuations. Thus the {\em classical}, linear meson-skyrmion
interactions behave as
\eq
     \cL \sim \phi ~\!(\dot{A} ~{\rm terms})
\label{naive}
\en
if the Skyrme equations of motion are imposed.
The time derivatives in eq.~\ref{naive}
imply a suppression of at least a factor of
$\Nc^{-1}$ compared to the behaviour given
in~\ref{Ndepend}.
If the chiral fluctuations are only considered classically,
this would be the end of the story. However, the classical
argument cannot be trusted since the chiral
fluctuations must be quantized in order to properly account for
the mesons as quantum fields.\cite{hjs2,hjs3}    This is the origin
of a number of complications.

The canonical momenta of the rotating skyrmion and associated
mesons are not independent, and the separation into components
transverse to the zero-mode solutions is ``gauge dependent.''
Correctly calculating an $S$-matrix element---which must
be gauge invariant---such as the $\pi$-N scattering amplitude
is a complicated task.\cite{hayashi1}
The advance made by the authors of ref.~\mite{hayashi2}
is to calculate  the Born terms of
$\pi$-N scattering in a manifestly
gauge-invariant manner in the (two-flavour) Skyrme
model. They obtain the same result as that given by
the gauge-dependent
separation of skyrmion and meson momenta,\cite{hayashi1} confirming that
the Yukawa coupling
$\gcup$ has the large $\Nc$ behaviour of eq.~\ref{Ndepend}.
This suggests that the large-$\Nc$ analysis of CP conserving $\pi$-N
scattering in the Skyrme model
is consistent with the implications of
chiral symmetry.

The gauge-invariant calculation presented in ref.~\mite{hayashi2}
is considerably simpler than the calculations in a gauge
that separates the skyrmion and meson momenta. As a result,
application of this method to the calculation
of the $\pi-N$ couplings in the three-flavour Skyrme model yields
a straightforward estimate of the large-$\Nc$ behaviour of
$\gcup$ and $\bgcup$.

\vspace{1.0cm}
\noindent {\bf 2. The CP Violating Skyrmion}
\vspace{0.3cm}
\onward
\setcounter{footnote}{1}

In order to obtain strong CP violation, one must consider
a ${\rm U}(3)$ non-linear $\sigma$-model
so as to include ~({\it i}) an effective interaction representing
the ${\rm U}(1)$ anomaly and ~({\it ii}) a CP violating mass-matrix.
The appropriate effective lagrangian in the large $\Nc$ limit is
then\cite{cpviol}
$$
  \cL (U) = \cL_{S} (U) + {f_{\pi}^2 \over 16}
 \Tr ( M U + M^{\dagger} U^{\dagger} - M - M^{\dagger})
+ {a f_{\pi}^2 \over 64 \Nc} (\Tr [\ln U - \ln U^{\dagger}])^2
 + \cL_{WZ}
$$
\eq
\label{effL2}
\en
where $\cL_{WZ}$ is the Wess-Zumino $5$-form,
$U$ is a $\u3$ matrix field,
$a$ depends on the physical meson masses,
\eq
  \cL_{S}(U) = {f_{\pi}^2\over 16}
           \Tr [ (\partial_{\mu} U)(\partial_{\mu}U^{\dagger})]
    + {1\over 32} e^{-2} \Tr \{ [(\partial_{\mu} U)U^{\dagger},
                                 (\partial_{\nu} U) U^{\dagger}]^2 \}
\en
and
\eq
  M= {\rm diag}[ M_1 \exp(i {\beta_1 \over M_1}),
                 M_2 \exp(i {\beta_2 \over M_2}),
                 M_3 \exp(i \epsilon {\beta_3 \over M_3}) ] ~.
\en
In order to prevent strong CP violation from creating
Goldstone bosons from the vacuum we require\cite{dnchiral,dixon}
\eq
       \begin{array}{l}
         \beta_1 = \beta_2 = \beta_3 \equiv \beta  \\[0.4cm]
     \ds    \epsilon = { a + M_3 \over a + M_1}
    \end{array}
\en
up to terms of $\cO(\beta^2)$.
It is convenient to define $\theta = {\rm arg~det} M$, the
effective CP angle, and to set
\eq
 \theta =    \lambda_{\theta}
\left( {1 \over M_1} +{1 \over M_2} + {\epsilon \over M_3} \right)
\en
so that $\lambda_\theta = \beta$.

The classical $\su3$ Skyrme ansatz, neglecting CP violation,
is\cite{anw,su3}
\eq
    U_{0}(\bfvec{r}) =
      \left(  \begin{array}{cc}
                    \cos F + i \bftau \cdot \bfvec{r} \sin F &
                          \begin{array}{c}
                                0 \\
                                 0
                           \end{array}   \\
                \begin{array}{cc}
                           0 & 0
                      \end{array}
             &          1
             \end{array}      \right)
  =  {\sqrt{2} \over 3} (\cos F -1) \lambda_8 +
    {1 \over 3}   (2\cos F + 1) I + i \bftau \cdot \bfvec{r} \sin F
\label{clskyrm}
\en
where $F= F(r)$.
As emphasized by the authors of ref.~\mite{dixon} the
CP violating term in $\cL$ reacts back on the skyrmion. We adopt
their generalization of ansatz~\ref{clskyrm} and consider
\eq
  \begin{array}{ccl}
   U_c (\bfvec{r}) & = &
\ds \exp( {2i\over f_{\pi}} [ \sqrt{{2\over 3}} (f_0 (r) - \delta) I
                   + \lambda_8 f_8 (r) ])  ~U_0 (\bfvec{r}~\!)
  \end{array}
\label{genskyrm}
\en
where
$$
\delta =  {\sqrt{6} \over 4} {\lambda_\theta f_\pi \over a + m_\pi^2}
$$
is the further vacuum shift required in order to prevent
the creation of Goldstone bosons from the vacuum.\cite{dixon}
The ansatz in eq.~\ref{genskyrm} remains
spherically symmetric under combined isospin and spatial rotations.
Neglecting isospin violation one can take
\eq
     \begin{array}{ccc}
  M_1 =  M_2  = m_{\pi}^2 &
   ~~~M_3 = 2 m_K^2 - m_\pi^2~~~ &
3a/\Nc = m_\eta^2 + m_{\eta^\prime}^2 - 2 m_K^2
   \end{array}
\en
so that $\lambda_\theta \sim \shalf m_\pi^2 \theta$.
Since it does not affect
any issues of principle, we neglect $\eta$-$\eta^\prime$ mixing
and corrections of order $\cO (M_1/a)$
in order to simplify the discussion. Taking these effects
into account would only produce small {\em numerical} corrections of
our results.
The CP violating ($CPV$) part of the lagrangian, which
contains contributions from both the mass and anomaly terms,\cite{dixon}
provides a source for $f_0(r)$ and $f_8(r)$ :
\eq
    \delta \cL_{CPV} =
   { \lambda_{\theta} f_\pi \over 2 \sqrt{3}}
 (1- \cos F) (f_8 + \sqrt{2} f_0) +  \cO ( {M_1 \over a}) ~.
\en
The function $F(r)$ is unchanged to order $\cO(\theta)$ by the
presence of this perturbation.

Neglecting terms of order $\cO ( m_\pi^2 / m_\eta^2)$
one can use this source to solve for $f_0$ and $f_8$ in terms of $F(r)$.
The result is\cite{dixon}
\eq
   \begin{array}{ccc}
    f_8 (r) &  =  & \ds { \lambda_{\theta} f_\pi \over 2 \sqrt{3} }
       \int { {\rm d}^3 r^\prime \over 4\pi}
  {e^{-m| \bfvec{r} - \bfvec{r}^\prime|} \over  | \bfvec{r} -
                               \bfvec{r}^\prime|}
  [ 1 - \cos F(r^\prime)]      \\[0.4cm]
    f_0 (r) &  =  & \ds
           { \sqrt{2} \lambda_{\theta} f_\pi \over 2 \sqrt{3} }
     \int  { {\rm d}^3 r^\prime \over 4\pi}
{e^{-m^\prime| \bfvec{r} - \bfvec{r}^\prime|} \over |\bfvec{r} -
                               \bfvec{r}^\prime|}
  [ 1 - \cos F(r^\prime)]
   \end{array}
\label{source}
\en
where we have neglected $\eta$--$\eta^\prime$ mixing, and
have written $m= m_\eta$ and $m^\prime = m_{\eta^\prime}$.
The angular integration in eqs.~\ref{source} gives
\eq
   \begin{array}{ccl}
    f_8 (r) &  =  &\ds  { \lambda_{\theta} f_\pi \over 2 \sqrt{3} m}
       \int_0^\infty {\rm d}r^\prime
      G_m(r,r^\prime) [ 1 - \cos F(r^\prime)]      \\[0.4cm]
f_0 (r) &  =  &\ds
  { \sqrt{2} \lambda_{\theta} f_\pi \over 2 \sqrt{3}m^\prime}
     \int_0^\infty  {\rm d}r^\prime
      G_{m^\prime}(r,r^\prime) [ 1 - \cos F(r^\prime)]
   \end{array}
\label{sourcesol}
\en
where
\eq
 G_{m} ( r, r^\prime) =
 \left\{  \begin{array}{l}
\ds \left({r^\prime \over r} \right) e^{-m r^\prime} \sinh(mr)  ~~ r < r^\prime
\\[0.5cm]
\ds \left({r^\prime \over r} \right) e^{-m r} \sinh(mr^\prime)  ~~ r > r^\prime
\end{array}   \right.
\en

\noindent The contributions of $f_8(r)$
and $f_0(r)$ will be  important in what follows.

\vspace{1.0cm}
\noindent {\bf 3. Meson-Nucleon Scattering in the Skyrme Model}
\vspace{0.3cm}
\onward
\setcounter{footnote}{1}

Here we generalize the analysis of ref.~\mite{hayashi2} to
the CP-violating, three-flavour model described in the
previous section. Denote the meson field (its quantum nature
signaled by the absence of the subscript $c$) by
\eq
U(\bfvec{r},t) = \lambda_\alpha U_\alpha  ~~~~\alpha= 0, 1,\ldots, 8 ~.
\label{mesf}
\en
It is convenient to write
\eq
    \begin{array}{ccc}
      U_\alpha & = &  s_\alpha  +i p_\alpha
 \end{array}
\en
where $s_\alpha$, and $p_\alpha$ are {\em real} fields. The
non-linear $\sigma$ model constraint that $U$ is a $\u3$ field,
\eq
   U^\dagger U = I ~,
\en
gives nine constraint equations.  Therefore we can regard
the $s_\alpha$ fields as dependent variables expressible in terms
of the dynamical fields
$p_\alpha$. (These nine independent real fields
embody the pseudoscalar nonet.)

Substituting eq.~\ref{mesf} into the lagrangian (eq.~\ref{effL2})
we obtain (exhibiting all time derivatives explicitly)
\eq
 \begin{array}{ccc}
  \cL & = & \ds
     \half \dot{p}_\alpha K_{\alpha \beta} \dot{p}_{\beta}
        + W_\alpha \dot{p}_\alpha  -  \cV(p,\nabla p)
 \end{array}
\label{insub}
\en
where $\alpha, \beta = 0, 1, \ldots, 8$ and $W_\alpha$ is obtained
from the Wess-Zumino term.
(This is the analogue of eq.~3.3 of ref.~\mite{hayashi2}.)
In the weak field limit
\eq
   K_{\alpha \beta} = {f_\pi^2 \over 4} \delta_{\alpha \beta} + \ldots
\en
Beyond this, the detailed forms of $K_{\alpha \beta}$, $W_{\alpha}$,
and $\cV(p, \nabla p)$ will not be needed.
The canonical momenta are
\eq
     \Pi_\alpha = {\delta L \over \delta \dot{p}_\alpha} =
             K_{\alpha \beta} \dot{p}_\beta + W_\alpha
\en
so that
\eq
    \dot{p}_\alpha = (K^{-1})_{\alpha\beta} (\Pi_\beta - W_\beta) ~~.
\en
Consequently, the Hamiltonian is, up to terms of $\cO (\hbar^2)$ since we
have ignored the problem of operator ordering,
\eq
    \cH= {1\over 2} \int {\rm d}^3 x
     (\Pi - W)_\alpha (K^{-1})_{\alpha\beta} (\Pi - W)_\beta
          + \cV (p,\nabla p)
\label{ham}
\en
and the canonical commutation relations are
\eq
    [  p_\alpha (x) , \Pi_\beta (y) ] \delta(x^0 - y^0) =
              i \delta_{\alpha \beta} \delta^{(4)}(x-y) ~.
\label{ccr}
\en
Due to the derivative couplings
$K_{\alpha\beta} \neq {f_\pi^2\over 4} \delta_{\alpha\beta}$ and
$\cH_{{\rm int}} \neq - \cL_{{\rm int}}$, as
is evident from eq.~\ref{insub}-\ref{ccr}.
This is the essential reason that the classical argument
for the absence of Yukawa couplings described in the introduction
fails.

The skyrmion soliton is the {\em static} but {\em rotated}
solution (\ie, $U^s = A U_c A^{-1}$) to the equations of motion
obtained from eq.~\ref{insub}.
(The superscript $s$ will denote the static, classical part of a
quantity in the following.) In terms of
the field of eq.~\ref{genskyrm}, this solution is given by
\eq
     p^{s}_\alpha = \half {\rm Im} ~\!\Tr (\lambda_\alpha A U_c A^{-1})
\label{psrot}
\en
where $A \in \su3$ rotates the ansatz $U_c$ to an
arbitrary direction in $\su3$ space and for the static
solution we neglect any time dependence of $A$.

In order to obtain the S-matrix for the meson-nucleon elastic scattering,
we will analyze the LSZ reduction formula.
To carry this out we must identify
the interpolating field $\Phi_\alpha$.
In fact, it is proportional to
the quantum field $p_\alpha$ in eq.~\ref{insub} (not  the
classical $p_\alpha^{s}$).
Since
$\ds \lim_{r \rightarrow \infty} U_c(r) = {\rm I}$, we can
pick out the proportionality constant from the non-linear meson fields
which appear in eqs.~\ref{matf} \&~\ref{Linsert} by using the
fact that
$\ds \lim_{r \rightarrow \infty} U(\bfvec{r},t) = U_{2\phi}$:
\eq
     U_{2\phi} \sim 1 + {2i \over f_\pi} \bfvec{\lambda}\cdot \bfvec{\phi}
               + \cO({\phi^2\over f_\pi^2})
\en
so that
\eq
              \Phi_\alpha = {f_\pi \over 2} p_\alpha  ~.
\label{intf}
\en
Therefore
\eq
          \Phi_\alpha = {f_\pi \over 4} {\rm Im} \Tr
            [\lambda_\alpha U(\bfvec{r},t)]
\label{propto}
\en
where $U(\bfvec{r},t)$ is the complete field
$U_\phi A U_c A^{-1} U_\phi$ that  enters $\cL(U)$.

The reduction formula for our problem is given in terms of the
source term $\cJ(x)$, $\Phi$ and $\dot{\Phi}$, as follows
\eq
   \begin{array}{l}
\ds    S_{fi} =  \delta_{{\rm f i}} +  i^2 \int {\rm d}^3
                 \bfvec{r}^\prime {\rm d} t^\prime
   \int {\rm d}^3  \bfvec{r} {\rm d}t
f^*_{k^\prime} (x^\prime) f_k(x)  \\[0.3cm]
{}~\times ~\VEV{ B^\prime(\bfvec{p}^\prime)
               | T(\cJ_\beta(x^\prime) \cJ_\alpha(x))
      + \delta (t^\prime-t)[\dot{\Phi}_\beta (x^\prime), \cJ_\alpha (x) ]
       - i \omega_{k^\prime} \delta (t^\prime-t)
             [ \Phi_\beta (x^\prime) , \cJ_\alpha (x) ] | B(\bfvec{p})} \\
        ~
  \end{array}
\label{redform}
\en
with $f_{\bfvec{k}}(x) = e^{-i k\cdot x}/\sqrt{(2\pi)^3 2 \omega_k}$,
where $k\cdot x = \omega_k t - \bfvec{k}\cdot \bfvec{r}$.
The source term of the meson fields is
\eq
 \cJ_\alpha (x) = - [ H, [ H, \Phi_\alpha(x)]] + (-\nabla^2 + m_\pi^2)
     \Phi_\alpha(x)
\label{mesour}
\en
with the commutator computed using eq.~\ref{ham},~\ref{ccr}, and~\ref{intf}.
The result is
\eq
   \begin{array}{l}
 \ds  [H,[H,p_\alpha(y)]]  =  (K^{-1})_{\alpha \beta}
  {\delta   \cV (\bfvec{p} , \bfvec{\nabla} p )\over \delta p_\beta}  \\[0.5cm]
+ \half (\Pi-W)_\gamma
                (K^{-1}_{\alpha \beta} K^{-1}_{\gamma \delta , \beta}
             - 2 K^{-1}_{\gamma \beta} K^{-1}_{\alpha \delta , \beta})
        (\Pi-W)_\delta  \\[0.5cm]
     + (\Pi-W)_\gamma K^{-1}_{\alpha \beta} K^{-1}_{\gamma \delta}
     (W_{\beta, \delta} - W_{\delta, \beta})
  \end{array}
\label{comres}
\en
where for any quantity $T$ and any index $\alpha$, the notation
$T_{,\alpha}$ denotes $\delta T / \delta p_\alpha$.
We have not kept track of operator
orderings, so that eq.~\ref{comres} is accurate only up
to terms of $\cO(\hbar^2)$.
We again see that the presence of derivative couplings in $\cL$ and $\cH$
produces a different result in eq.~\ref{comres} than
that expected from classical considerations.

Similarly, using eq.~\ref{mesour} \&~\ref{comres},
we find that
\eq
   \begin{array}{ccl}
     [ \dot{p}_\beta(x), \cJ_\alpha(y)] \delta(x^0-y^0) & = &
\ds - i (K^{-1})_{\beta\gamma} {\delta \cJ_\alpha (y) \over \delta p_\gamma
(x)}
                           \delta(x^0-y^0)  \\[0.65cm]
   &  &  \ds + [ (K^{-1})_{\beta \gamma} , \cJ_\alpha (y)] (\Pi - W)_\gamma
   \delta (x^0 - y^0) \\[0.65cm]
   &   &\ds  - (K^{-1})_{\beta\gamma} [W_\gamma (x) , \cJ_\alpha (y) ]
                     \delta (x^0 - y^0)
\end{array}
\en
with
\eq
         [ \dot{\Phi}_\beta(x), \cJ_\alpha(y)] \delta(x^0-y^0)
   =  {f_\pi \over 2} [ \dot{p}_\beta(x), \cJ_\alpha(y)] \delta(x^0-y^0)
\label{alsneed}
\en
We also need
\eq
      [ \Phi_\beta(x), \cJ_\alpha(y)] \delta(x^0-y^0)
 =  i {f_\pi \over 2} {\delta \cJ_\alpha (y) \over \delta \Pi_\beta (x)}
              \delta(x^0-y^0) ~.
\label{alneed}
\en
The above equations are again accurate only to order $\hbar$.

In order to evaluate the Born terms from eq.~\ref{redform}
we need certain one-baryon matrix
elements, which will be analyzed in terms of the
$1/\Nc$ expansion.

The {\em crucial} point is that in the matrix elements of operators
taken between single-baryon states, one may replace the field operators
with the classical fields evaluated with the skyrmion solution
(\ref{psrot}), \ie, $p_\alpha(x) \rightarrow p^s_\alpha (x) $.
This idea goes all the way back to reference~\mite{onesol}.
The reason is that the fluctuating parts
vanish since the ``in'' and ``out'' meson states form a Fock space
on the one-baryon subspace. As emphasized in ref.~\mite{hayashi2}
it is not necessary to explicitly seperate $\Phi_\alpha$ into
static and fluctuating parts, since the one-baryon meson matrix
elements will automatically pick out the static solution.

In order to arrange the one-baryon matrix elements in powers of
$1/\Nc$, we need to order the various classical field quantities
in terms of  powers of $\Nc$. By the classical equations of motion
\eq
    {\delta \cV (p^s , \nabla p^s) \over \delta p^{s}_\alpha}= 0
\en
Note that the ordering in $1/\Nc$ is the {\em same} for the CP
conserving (CPC) and CP violating (CPV) terms in $p_\alpha^s$,
to leading order in CP violation, as can be seen from the following.
{}From eq.~\ref{psrot}, the CP conserving term in $p_\alpha^s$ is
\eq
    (p^s_\alpha)_{{\rm CPC}} = \half \tr (\lambda_\alpha A
                   {\bftau}
             A^{-1})\cdot \bfvec{r} \sin F(r)
\en
so that
\eq
   (p^s_\alpha)_{{\rm CPC}} \sim \cO (1)
\label{firord}
\en
in the $1/\Nc$ expansion.
Then,
\eq
    \begin{array}{ccl}
      \ds  p^s_\alpha &  = &
     (p^s_\alpha )_{{\rm CPC}}
       +   (p^s_\alpha )_{{\rm CPV}} \\[0.4cm]
     &  \sim & \ds \cO (1) [ 1 + \lambda_\theta/m_\eta^2 ] \\[0.4cm]
    &  \sim & \ds \cO (1) [ 1 + ({m_\pi^2 \over m_\eta^2} ) \theta]
   \end{array}
\label{bord}
\en
where we have used eqs.~\ref{genskyrm} and \ref{sourcesol}.
Therefore, the $1/\Nc$ estimates that follow hold
in both the CP violating and conserving cases.

The estimates in eqs.~\ref{firord} and~\ref{bord},
 and eq.~\ref{ccr} imply that
\eq
   (\Pi_s) \sim \cO (1)
\label{ordPi}
\en
Since
\eq
    (K_s)_{\alpha \beta} \sim \cO (f_\pi^2) \sim \cO (\Nc)
\en
one also has
\eq
    \begin{array}{lcl}
        \dot{p}^s_\alpha & \sim & \cO (\Nc^{-1}) \\[0.25cm]
        (K^{-1})^s_{\alpha \beta} & \sim & \cO (\Nc^{-1}) \\[0.25cm]
        (W^s)_\alpha & \sim & \cO (1)
     \end{array}
\label{ordrest}
\en
for these classical quantities.
With these ingredients we can now make
$1/\Nc$ estimates of the one-baryon matrix elements.

For the first term in eq.~\ref{redform} we need
\eq
   \begin{array}{ccl}
  \vev{ B | \cJ_\alpha (x) | B^\prime} &  = &
   \ds \vev{ B | (-\nabla^2 + m_\pi^2) \Phi_\alpha +
                  \partial_t^2 \Phi_\alpha | B^\prime} \\[0.4cm]
 & = & \ds {f_\pi \over 2} \{ \vev{ B |  (-\nabla^2 + m_\pi^2) (p_s)_\alpha
             | B^\prime} + \cO(1/\Nc^2) \}
  \end{array}
\label{matel}
\en
which holds because
all the terms in eq.~\ref{comres}, evaluated with the classical
solution, contribute at most to the order $\cO(1/\Nc^2)$
correction.

Similarly, from eq.~\ref{alneed}
\eq
   \begin{array}{ccl}
          \vev{B | [ \Phi_\beta (x) , \cJ_\alpha (y) ] \delta (x^0 - y^0)
                 | B^\prime} & = &
      \ds     i {f_\pi \over 2} \vev{ B | \left[ {\delta \cJ_\alpha (y) \over
                           \delta \Pi_\beta (x) } \right]_s | B^\prime}
                 \delta (x^0-y^0)  \\[0.4cm]
      & \sim & \cO (f_\pi^{-3}) \sim \cO (\Nc^{-{3\over 2}})
  \end{array}
\en
and from eq.~\ref{alsneed}
\eq
     \begin{array}{ccl}
\vev{B | [ \dot{\Phi}_\beta (x) , \cJ_\alpha (y) ]
  \delta (x^0 - y^0)| B^\prime} & = &
  \ds   {f_\pi \over 2} \{-i\vev{ B | (K_s^{-1})_{\beta\gamma}
     \left[ {\delta \cJ_\alpha (y) \over
                           \delta p_\gamma (x) } \right]_s | B^\prime}
                 \delta (x^0-y^0)  + \cO (1/\Nc^2)\} \\[0.4cm]
      & \sim & \cO ( 1 )
  \end{array}
\label{issone}
\en
Therefore only eq.~\ref{matel} and~\ref{issone} contribute to the Born terms
in the large $\Nc$ limit.
In addition, only equation~\ref{matel} appears in the
pole term, and so gives both the CP conserving and CP violating
Yukawa couplings, while  eq.~\ref{issone} contributes to a non-pole
contact term.

\vspace{1.0cm}
\noindent {\bf 4. The $\pi$-N Coupling Constants}
\vspace{0.3cm}
\onward
\setcounter{footnote}{1}

The classical, CP-violating  solution
is, from eqs.~\ref{genskyrm}, \ref{intf},
and~\ref{propto},
\eq
   (\Phi^s_\alpha)_{{\rm CPV} } = {1\over 30 \sqrt{3}}
\left\{ [ \sqrt{2} (f_0(r) - \delta) + f_8(r) ]
 ( \cos F(r) - 1) + 4 f_8(r) \right\} \Tr ( \lambda_\alpha A \lambda_8 A^{-1})
\label{cpfield}
\en
Therefore, to leading order in $1/\Nc$,
\eq
    \begin{array}{rcl}
\bgcup (\bfvec{k}) & = &
\ds  \int {{\rm d}^3 r \over (2\pi)^3} ~\! e^{-i \bfvec{k}\cdot
\bfvec{r}}
\vev{B| (\cJ_\alpha)_{{\rm CPV}} | B^\prime} \\[0.5cm]
 & \simeq &  \ds  \int {{\rm d}^3 r \over (2\pi)^3} ~\! e^{-i \bfvec{k}\cdot
\bfvec{r}}
\vev{ B | (-\nabla^2 + m_\pi^2)(\Phi^s_\alpha)_{{\rm CPV}} | B^\prime}
 \end{array}
\label{fres}
\en
If the physical (mass-shell) coupling
\eq
    \bgcup = \bgcup (\bfvec{k})|_{\bfvec{k}^2 = - m^2_\pi}
\en
is to have a contribution from eq.~\ref{fres},  then
$(\Phi_\alpha^s)_{CPV}$ must have a pion
tail (\ie it must behave as
$\exp(-m_\pi r)/r$ as $r \rightarrow \infty$).
Since
\eq
  F(r)    \longrightarrow  \ds   C    {e^{-m_\pi r} \over r}
                    ~~{\rm as}~ r \rightarrow \infty
\en
it is clear that the $\cos F(r)$ factor does not have a  pion tail.
Further, eq.~\ref{source} shows that $f_8(r)$ and $f_0(r)$ have
(momentum space) poles at $m_\eta^2$ and $m^2_{\eta^\prime}$,
respectively, rather than at $m^2_\pi$, so that
the required pion tail is entirely absent from
eq.~\ref{cpfield}. By contrast, the CP conserving
coupling is given by (the analysis is
essentially the same as that
already discussed in ref.~\mite{hayashi2})
\eq
    \gcup (\bfvec{k}) = {2 \over 3} M_N f_\pi
         {\bfvec{k}^2 + m_\pi^2 \over |\bfvec{k}| }
     \int {\rm d}^3r j_1(k r) \sin F(r)
\en
where $ j_1(y)$ is the spherical Bessel function. Since
$\sin F(r)$ ($\sim F(r)$ as $r \rightarrow \infty$)
{\em does} have the required pion tail, $\gcup$ is non-vanishing
on-shell.

Therefore we conclude that $\bgcup$ is suppressed relative
to eq.~\ref{Ndepend}
by at least a factor of $\Nc^{-1}$, while the behaviour of
$\gcup$ is correctly given
by~\ref{Ndepend}.\footnote{Although the order $\cO (N^{\half})$
contribution to $\bgcup$ does not vanish off-shell, it is
suppressed by the factor $m^2_\pi /m^2_\eta$, as is evident
from eq.~\ref{bord}.}
It is important to note that since $\gcup$ is not
suppressed by factors of $\Nc^{-1}$, the stability of the linear
fluctuations about the classical solution cannot be the reason for
the suppression of $\bgcup$. The correct explanation rests with the
asymptotic behaviour of the skyrmion profile function eq.~\ref{clskyrm},
where one should compare
the coefficient of $\lambda_8$ with that of $\bftau$.

\vspace{1.0cm}
\noindent {\bf 5. Conclusions}
\vspace{0.3cm}
\onward
\setcounter{footnote}{1}

Our calculation shows that the
Yukawa couplings in the Skyrme model involve issues more
subtle than the vanishing of linear fluctuations about the
classical solution. The CP conserving coupling $\gcup$
is not suppressed, while the CP violating coupling is.
Although this supports the final result of
ref.~\mite{dixon}, we find that neglect of the pion
contribution to $\Dn$ in the large $\Nc$ limit
requires more analysis than indicated there
before one can be assured that the
direct contribution to $\Dn$ dominates.
The behaviour of
the CP violating coupling constant $\bgcup$ is also interesting
for other applications.\cite{new}

The primary value of the Skyrme model is as a phenomenological
approximation to the low-energy effective lagrangian of
QCD; therefore, one hopes that it substantially embodies the physics of
chiral symmetry and the large-$\Nc$ properties of QCD.
In fact, we have found that
simple QCD estimates based  on na\"{\i}ve quark-line
counting (\ie estimates similar to those of ref.~\mite{largeN}
but using the $\theta g^2 F \widetilde{F}$ interaction) agree
with our  Skyrme model results.
In regard to this, it would
have been peculiar if the Skyrme model, which does give the correct
large $\Nc$ behaviour for
the multi-meson skyrmion, did not, in the end,
give the correct Yukawa couplings as well.

\end{document}